\documentclass[epj]{svjour}

\usepackage{amsmath}
\usepackage{sidecap}
\sidecaptionvpos{figure}{t}

\RequirePackage[colorlinks,citecolor=blue,urlcolor=blue,linkcolor=blue]{hyperref}

\usepackage{graphics}
\usepackage{graphicx}
\usepackage{dcolumn}
\usepackage{bm}
\usepackage{color}
\usepackage{siunitx}
\usepackage{comment}
\usepackage{caption}
\usepackage{subcaption}
\usepackage{adjustbox}
\usepackage{float}
\usepackage{changepage}
\DeclareSIUnit[number-unit-product = {}]\sig{~\sigma}

\begin{document}
\title{Dechanneling Population at Extreme Crystal Bending with 6.5 TeV Proton Beam}
\author{Roberto Rossi\inst{1,2} \and Daniele Mirarchi\inst{1} \and Stefano Redaelli\inst{1} \and Walter Scandale\inst{1,2,3}}

\institute{ European Organization for Nuclear Research CERN, Geneva, Switzerland \and Blackett Laboratory, Imperial College, London SW7 2AZ, UK\and Instituto Nazionale Fisica Nucleare INFN, Sezione Roma I, Rome, Italy}
%

%
\abstract{
Beam measurements with bent crystals, installed in the Large Hadron Collider to assist multistage collimation system, provided information on hadron interactions with crystals in the multi--\unit{\tera\electronvolt} energy range.
In particular, the dechanneling population was observed through scans of deflected halo with collimators. 
Taking advantage of the fact that crystals with different values of curvature radii were present, the dependence of dechanneling on bending radius ($R$) was recorded. 
Dechanneling was found to be enhanced in crystals with smaller bending radius, because it is too close to the critical value $R_c$ at the LHC energy of \qty{6.5}{\tera\electronvolt} where channeling is lost. 
Data analysis and comparison to simulation results provided a better understanding of the phenomena and could be used to define specifications for more performing crystals in future upgrades of the crystal collimation system.
} 

\maketitle
\section{Introduction}
\label{sec:intro}

Charged particles entering in the highly pure crystalline lattice of a crystal may be captured in channeling states~\cite{lindhard}. If the crystal is bent, channeled particles may be deflected by an angle equal to the crystal bending angle $\theta_b$~\cite{tsyg,taratin}.
Applications of bent crystals for beam manipulations in particle accelerators have considerably increased the interest for a deeper understanding of crystal particle interactions.
Bent crystals were used to produce and split simultaneous kaon beams in the NA48 experiment at CERN~\cite{na48doble1996novel,na48fanti2007beam}. 
Crystal assisted proton extraction is exploited in the U70 synchrotron at IHEP in Protvino~\cite{afonin2001high,afonin2002investigations}. Loss reduction at the SPS  extraction relies on a bent crystal acting as a scatterer~\cite{velotti2019demonstration}.
An experimental set-up was installed in the CERN Large Hadron Collider (LHC) to study channeling with proton and heavy ion beams for halo collimation.
Channeling was observed for the first time at the unprecedented energy of \qty{6.5}{\tera\electronvolt}~\cite{lhc_ch,redaelli2021first}.
In addition to the fundamental purpose of studying the feasibility of a crystal collimation system~\cite{dan_layout,dan_th,rossi_phd}, this set-up offers the possibility of investigating other coherent phenomena at energies not accessible in any other facility.

In this paper, we discuss the observation of nuclear dechanneling in some of the past LHC crystal test runs. The interesting feature of our results relies on the fact that the crystal bending radius is close to the critical one because of the large energy of our test. The main consequence of this is that the dechanneling rate increases, whilst the planar channeling efficiency decreases.

\section{\label{sec:dechanneling}Planar Channeling and Critical Radius}

\begin{SCfigure}[][t]
\includegraphics[width=0.65\textwidth]{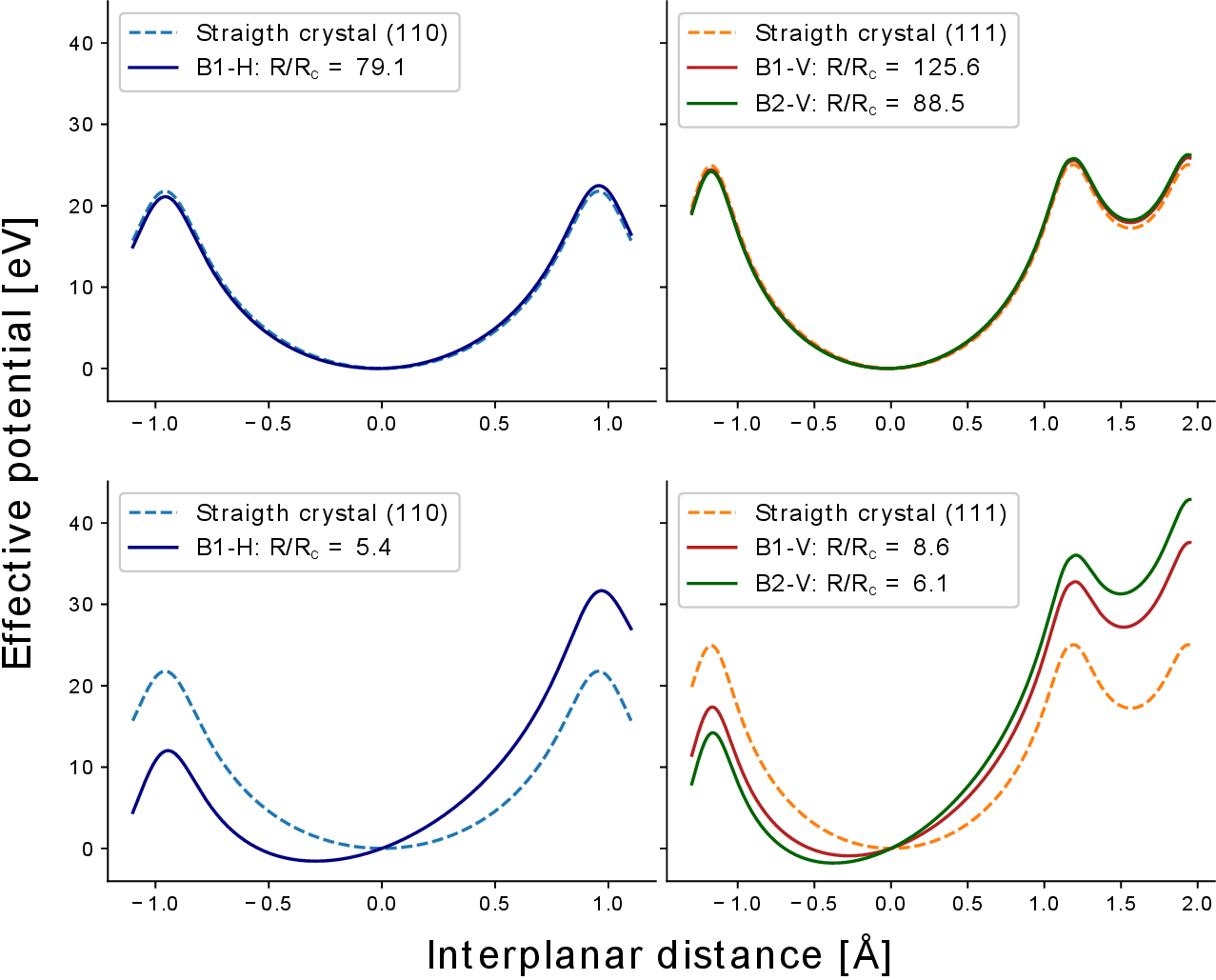}
  \caption{Crystal bending effect on the crystalline potential well for (110) (left column) and (111) (right column) crystals. Bending effect is shown in solid lines for crystals with the same parameters as the ones used for the LHC measurements. The top row shows the potentials at \qty{450}{\giga\electronvolt}, while the bottom row shows the ones at \qty{6.5}{\tera\electronvolt}. Dashed lines show in all cases the straight crystal potential as a reference.}
  \label{fig_potential}
\end{SCfigure}

For an optimal beam steering, bent crystals should have constant curvature.
The most efficient way to obtain it consists in applying a primary mechanical stress that generates a highly regular secondary deformation, via anti-clastic or Quasi-Mosaic effects. In both cases, the secondary curvature is almost circular.
The crystals used in LHC are of two types that differ for the shape, the deformation method and the bent crystalline planes. 
Strip crystals~\cite{Afonin,strip_lhc} are slim bars deformed by anti-clastic effect along the (110) planes, while the Quasi-Mosaic ones are plates deformed by the homonymous elastic reaction~\cite{qm_model} along the (111) planes.
The inter--planar potential well is also different because the (110) planes are equidistant and produce a quasi--parabolic potential that repeats identically after one inter--planar distance. 
Instead, the (111) planes exhibit a pattern with two interleaved inter-planar distances, one three times larger than the other and thus produce two interleaved potential wells, one about 5 times deeper than the other.
The curvature radius $R$ defines the deflection $\theta_b$ imposed to the channeled particles for a given crystal length, $l$, as: 
\begin{equation}\label{eq:bend}
\theta_b = l /R.
\end{equation}
\noindent The crystal bending causes an asymmetry of the inter--planar potential well in which particles are channeled.
In fact, the lower the distance from the centre of curvature (the center of the circle of radius $R$), the higher the nuclei density: the potential barrier is higher when moving toward the centre of curvature.
The channeling condition is fulfilled in bent crystals only if the lowest barrier is larger than the particles transversal energy.
Thus, a condition on the bending radius can be imposed by defining the critical radius $\mathrm{R_c}$ above which particles can no longer be trapped between crystalline planes. 
This value is better determined using the inter-planar electric field $U'(x)$:
\begin{equation}\label{eq:critrag}
R_c = \frac{pv}{U'(x_{\rm max})},
\end{equation}

\noindent where $x_{\rm max}$ is the largest amplitude of the particle oscillating trajectory. 
Coherent phenomena that appear in straight crystals are influenced by $R_c$ in bent crystals. 
In particular the channeling acceptance, defined by the critical angle $\theta_{\rm c} = \sqrt{2U_{\rm max}/pv}$, where $U_{\rm max}$ is the potential well, scales as $(1-R_c/R)$, whilst the dechanneling characteristic length ($L_D$), accounting for the e-folding decrease of the channeled intensity along the crystal length, scales as $(1-R_c/R)^2$~\cite{biryukov}.
Dechanneling might be caused by interaction with crystalline lattice nuclei or with free electrons in the crystalline channel.

The channeling efficiency depends on crystals characteristics (material, lattice purity,  bending) as well as on the particle energy.
The potential well for bent crystals, with the same crystal parameters and beam energy of the measurements presented in paper, are shown in Fig.~\ref{fig_potential}.
The effect of the larger $R_c$ value for the smaller potential well in the QM crystal is not considered in this paper, knowing that this channel only account for less than one third of the crystal volume and that no difference in channeling efficiency has been recorded in precedent publication.
Also, the value of $U'(x_{\rm max})$ are very similar for (110) crystal and the large channel in the (111); therefore a common value of $R_c$ is considered for the rest of the paper.

\section{\label{sec:layout}The LHC Crystal Collimation Experimental Layout}


\begin{SCfigure}[][t]
  \centering
  \includegraphics[width=0.65\textwidth]{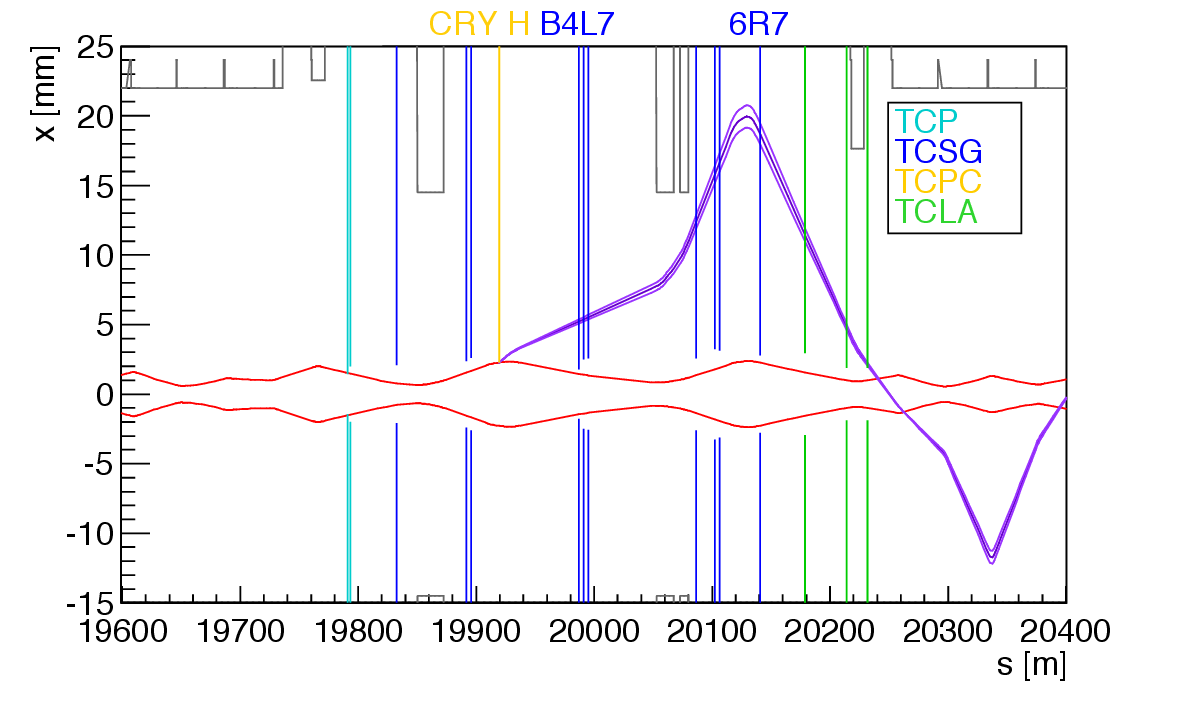}
  \caption{\protect\rule{0ex}{5ex}Trajectory of channeled halo particles (purple lines) at \SI{6.5}{\tera\electronvolt} for the horizontal crystal. The envelope of the circulating beam (red lines) at amplitudes of (\SI{5.5}{\sigma}), is defined by the crystal aperture . Vertical lines indicate the aperture of the IR7 collimators in the plane of interest.}
  \label{fig_layout}
\end{SCfigure}

A layout for crystal collimation studies has been conceived and optimized within the present LHC collimation system~\cite{dan_layout}. 
Crystal assemblies called TCPC, consisting in a silicon crystal mounted on a high-precision goniometer for angular adjustment~\cite{gonio2}, have been installed in the betatron collimation insertion of LHC~\cite{assman}.
The crystal bending angles and lengths were optimized through semi-analytical studies to:
\begin{itemize}
    \item intercept the channeled halo with enough clearance with downstream secondary collimators (TCSG);
    \item respect aperture constraints given by the beam pipe and the physical aperture of the elements downstream the crystals;
    \item maximize collimation cleaning performance.
\end{itemize}

\begin{table}[t]
\setlength\tabcolsep{3.8pt}
\centering
	\caption{Installation position, collimation plane and main optics parameters for the crystal devices (TCPC) and secondary collimators (TCSG) used as absorbers for the channeled halo. Nominal optics parameters are used. Longitudinal installation positions are measured from the first LHC collision point. The position of the middle point of the collimation insertion, IP7, is at \SI{19994.16}{\meter}. Note that B2 installations are specular with respect to IP7, therefore they have the same optic parameters as B1V installations.}
	\label{tab_param}
		\begin{tabular}{lccccccccccccc}
		\hline\noalign{\smallskip}
		\textbf{Name}  & \textbf{Beam} & \textbf{$s$} & \textbf{Collimation}     & \textbf{Length} & \textbf{Mat.} & \multicolumn{2}{c}{\textbf{Bending}} & \textbf{$\beta_{x}$} & \textbf{$\beta_{y}$} & \textbf{$\alpha_{x}$} & \textbf{$\alpha_{y}$} & \textbf{$D_{x}$} & \textbf{$D_{y}$}\\
	 &	& \textbf{[\si{\meter}]}&\textbf{Plane} &  \textbf{[\si{\milli\meter}]} & & \textbf{[\si{\micro\radian}]} &\textbf{Planes} &\textbf{[\si{\meter}]} &\textbf{[\si{\meter}]} &\textbf{[\si{\radian}]} &\textbf{[\si{\radian}]} &\textbf{[\si{\meter}]} &\textbf{[\si{\meter}]}\\
		\noalign{\smallskip}\hline\noalign{\smallskip}
        TCPCV.A6L7 &B1 &19843.62 &Ver. & 4   & Si  &40 &111& 30.5 & 281.1 & 0.24&-2.63 & 0.08 &-0.65\\
        TCPCH.A4L7 &B1 &19919.49 &Hor. & 4   & Si  &65 &110&342.1 &  64.9 &-2.05& 0.84 &-0.68 &-0.27\\
        TCSG.D4L7  &B1 &19917.24 &Ver. &1000 & CFC &-- & --&333.0 &  68.9 &-2.01& 0.90 &-0.67 &-0.29\\
        TCSG.B4L7  &B1 &19987.16 &Hor. &1000 & CFC &-- & --&139.8 & 131.0 & 1.43&-1.25 &-0.53 &-0.05\\
        TCSG.D4R7  &B2 &20071.08 &Ver. &1000 & CFC &-- & --&333.0 &  68.9 &-2.01& 0.90 &-0.67 &-0.29\\
        TCPCV.A6R7 &B2 &20145.20 &Ver. &4    &Si   &55 &111& 30.5 & 281.1 & 0.24&-2.63 & 0.08 &-0.65\\

\noalign{\smallskip}\hline
\end{tabular}
\end{table}

An example is shown in Fig.~\ref{fig_layout}, where the deflected halo trajectory is determined using the LHC optic functions and the transport matrix formalism.
There are also reported the position of the collimators present in the region near the crystal. 
The primary collimators (TCP) are not used during the crystal measurement. 
Only selected secondary collimators (TCSG) are used to intercept the halo deflected by the crystal, and the shower absorbers (TCLA) are used at nominal machine aperture during the measurements.
The relevant parameter for crystals and the secondary collimators, used as absorbers, are listed in Tab.~\ref{tab_param}, while a detailed presentation on the whole machine configuration is presented in~\cite{rossi_phd}.
The required specifications for crystal manufacturers were a bending angle of \SI{50}{\micro\radian} and a length of \SI{4}{\milli\meter}.

\begin{SCfigure}[][t]
  \includegraphics[width=0.65\textwidth]{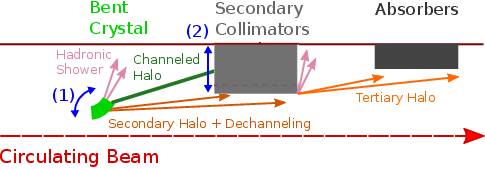}
  \caption{Illustrative view of the crystal angular scan (1) and of the collimator linear scan (2) that are used to demonstrate experimentally the onset of channeling. Angular scans are used to find the optimum channeling orientation when nuclear losses are minimized. Collimator scans probe the distribution of the channeled halo and the dechanneling region.}
  \label{fig_scheme}
\end{SCfigure}

The present LHC crystal collimation experimental setup~\cite{dan_th,dan_layout} is composed of four crystals: one for each plane and beam. 
Because of a problem during the alignment, and the impossibility to resolve it remotely, the horizontal crystal in the beam 2 (B2) is not considered for the studies of this paper~\cite{rossi_phd,rossi_note}.
In this experimental layout, it was possible to observe channeling in several LHC configurations, and also to characterize the crystal properties.
Losses at crystal and other positions are used to measure crystal characteristics in a circular accelerator.
Example of these measurements are shown in Fig.~\ref{fig_scheme}.

In channeling, particles are coherently deflected and separated from the beam core by the crystal. 
An evidence of the onset of channeling is the increase in losses at a downstream absorber location, compared with losses with the crystal oriented as an amorphous material. 
Moving the jaw of one of the absorbers toward the beam orbit enables studying the channeled halo.
By intercepting these particles with a linear transversal scan of a collimator jaw, the width and the aperture of the channeled halo are measured. 
Using a linear transport matrix formalism (as used in Fig.~\ref{fig_layout}), one can evaluate the crystal bending angle.

\begin{figure}[t]
  \begin{subfigure}[b]{0.51\textwidth}
  \centering
  \includegraphics[width=\textwidth]{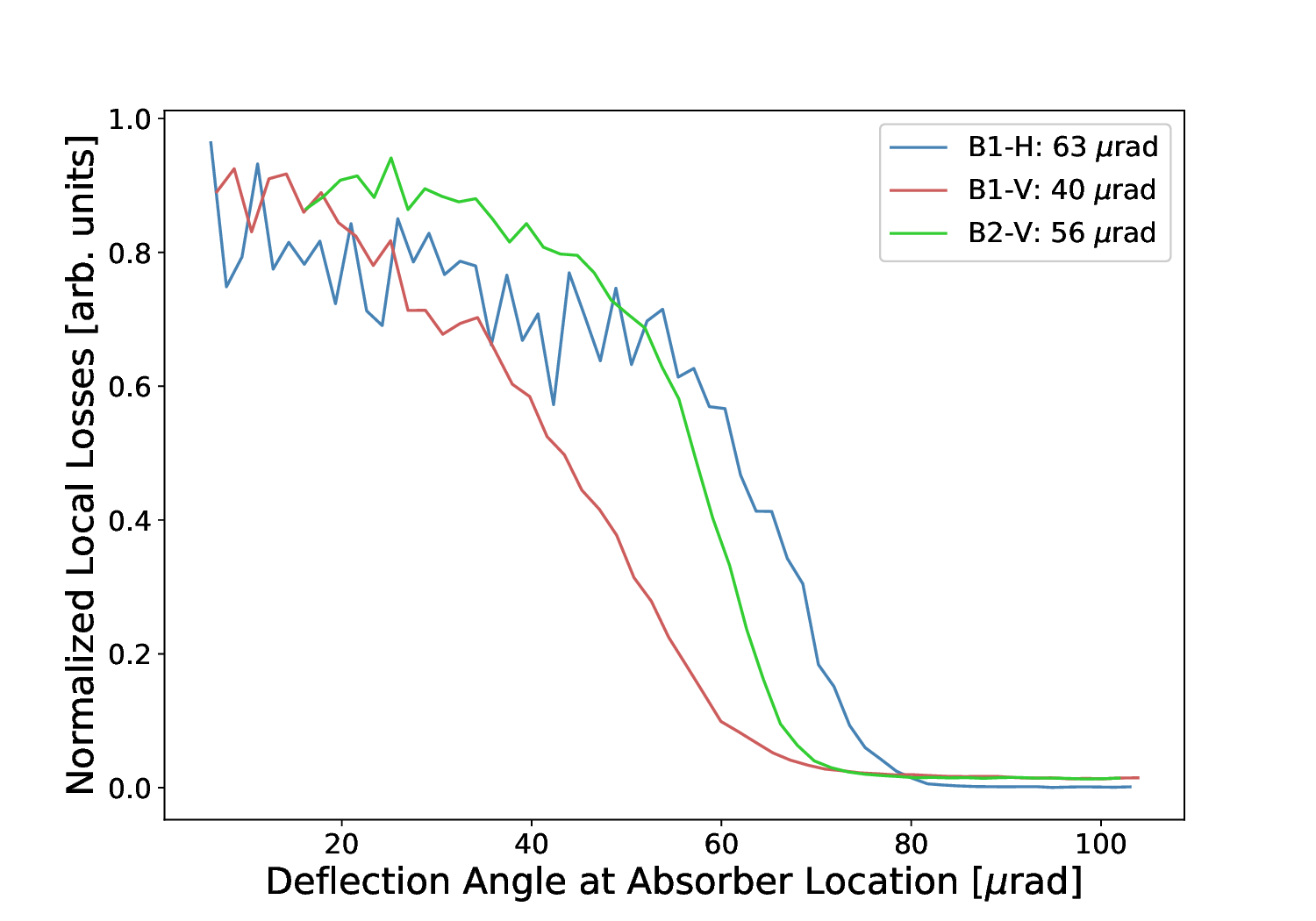}
  \caption{Injection energy measurements.}\label{sub:kick_inj}
  \end{subfigure}
  \begin{subfigure}[b]{0.51\textwidth}
  \centering
  \includegraphics[width=\textwidth]{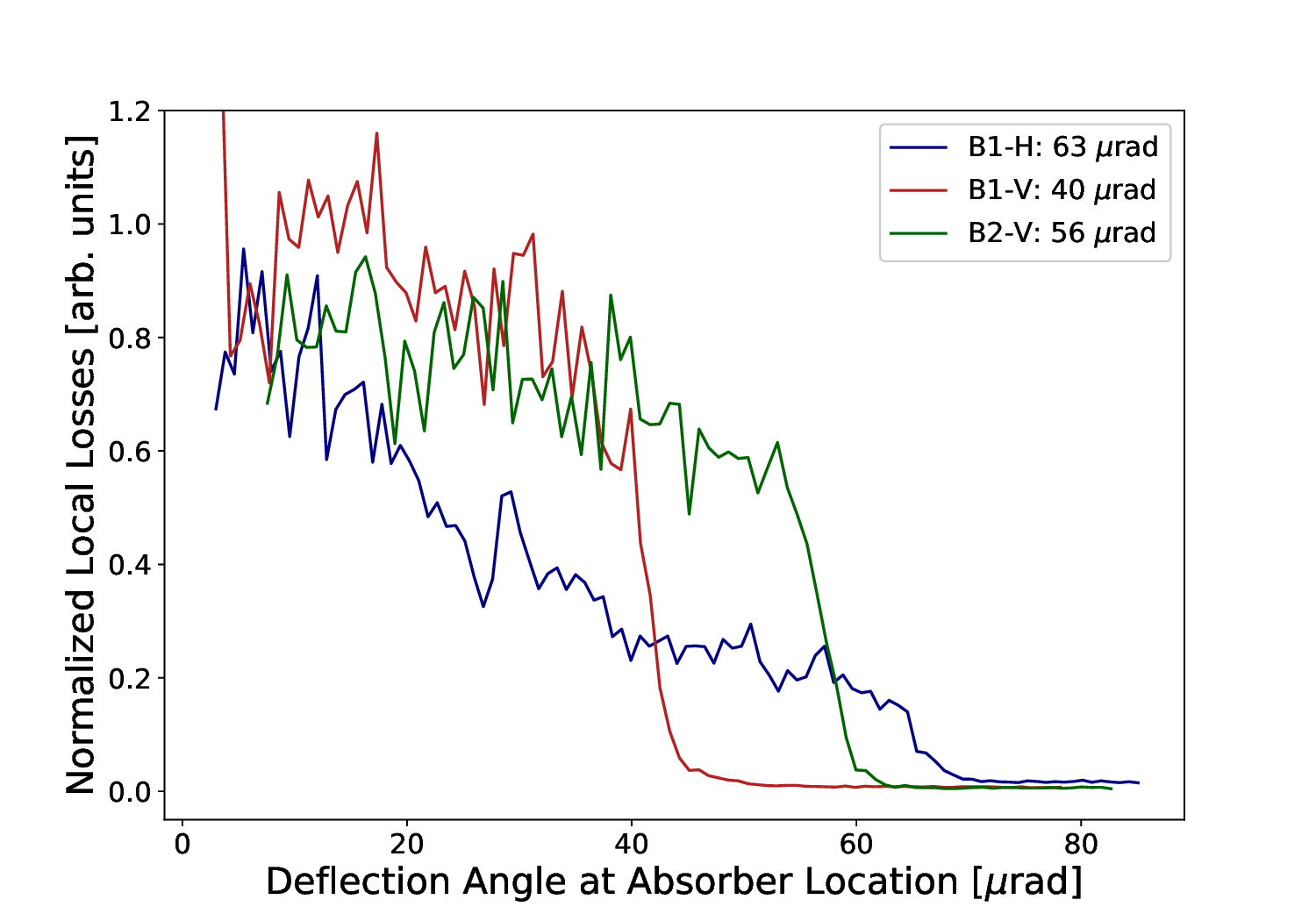}
  \caption{Top energy measurements.}\label{sub:kick_ft}
  \end{subfigure}
  \caption{Beam losses during collimator scan of the channeled beam. The BLM signal is normalized to beam flux and to the loss level at interception of the circulating beam. Data are plotted as a function of the equivalent deflection at collimator position, recorded at TCSGs during linear scans with the crystals in channeling orientation. A comparison between crystals behaviour is presented for injection~(\ref{sub:kick_inj}) and top energy~(\ref{sub:kick_ft}).}
  \label{fig_kick}
\end{figure}

\section{\label{sec:experiment}Experimental Results}

The collimator scans were performed with the crystal set as a primary obstacle\footnote{For these measurements, crystals were set at \SI{5.5}{\sigma}.} and with optimal angular orientation for channeling.
A downstream secondary collimator (TCSG) was used for the scans.
Tab.~\ref{tab_param} lists the layout names of these collimators with their orientations and relevant optic functions.

A selection of measurements are shown and compared in Fig.~\ref{fig_kick}, at injection (\SI{450}{\giga\electronvolt}, Fig.~\ref{sub:kick_inj}) and  top energy (\SI{6.5}{\tera\electronvolt}, Fig.~\ref{sub:kick_ft}).
Three crystals with different bending angles were considered.
The losses recorded downstream of the secondary collimators used for the scan are normalized to the beam flux and to the loss level when the circulating beam is touched.
On the abscissa the linear transport matrix formalism is used to evaluate the deflection angle at the absorber position:
\begin{equation}\label{eq_kick}
\theta_{\rm cry} = \frac{x_{\rm coll} - \sqrt{\frac{\beta_{\rm coll}}{\beta_{\rm cry}}}\left(\cos{\varphi}+\alpha_{\rm cry}\sin{\varphi}\right)x_{\rm cry}}{\sqrt{\beta_{\rm cry}\beta_{\rm coll}}\sin{\varphi}} - x'_{\rm cry},
\end{equation}

\noindent where $x$ and $x'$ are the transverse coordinates, $\varphi$ is the betatron phase advance between crystal and absorber and $\alpha$ and $\beta$ are the Twiss parameters. 
Eq.~\ref{eq_kick} is valid for both horizontal and vertical planes, and the labels ``cry'' and ``coll'' indicate the crystal and the collimator, respectively.
When the collimator jaw intercepts the channeled beam, a rise in losses is observed, that is proportional to the number of particles intercepted by the jaw; the full scan can be used to compute integral of the particles channeled by the crystals.
The following slow rising is instead proportional to the integral of the particles dechanneled from the crystals. 
These particles experience a lower deflection angles than the channeled ones.
When the collimator jaw touches the edge of the circulating beam, a spike is observed in the BLM signal, that is filtered in the analysis. 
Indeed, the spike suppression is applied to avoid problems to flux normalisation when the primary obstacle in the machine becomes the collimator jaw, as one can observe in Fig.~\ref{fig_kick}.

\begin{table}
\caption{Crystal bending angles and multiturn channeling efficiency computed from different data sets. Bending angles are computed as averages of measurements at both injection and top energy. Each crystal bending radius as a ratio to the critical one is reported at both energies. Finally, the multi--turn channeling efficiency averages are reported again per beam energy.}
\label{tab_summary}
\centering
\begin{tabular}{l c c c c c}
\hline\noalign{\smallskip}
		& Bending Angle          & \multicolumn{2}{c}{Bending Radius}& \multicolumn{2}{c}{Multi--turn Channeling}		\\
		& [\si{\micro\radian}]   & \multicolumn{2}{c}{[$R_c$]} & \multicolumn{2}{c}{Efficiency~[\si{\percent}]}           \\
Crystal	&   & Injection 		& Flat Top & Injection 		& Flat Top					 \\

\noalign{\smallskip}\hline\noalign{\smallskip}
B1-H	&\num{63.2\pm 1.7}	&\num{79.1\pm 2.1} &\num{5.4\pm 0.2}	&\num{71\pm 5}		&\num{27\pm 5}		\\
B1-V	&\num{39.8\pm 2.3}	&\num{125.6\pm 7.3} &\num{8.6\pm 0.5}	&\num{87\pm 5}		&\num{84\pm 5}		\\
B2-V    &\num{56.5\pm 1.8} 	&\num{88.5\pm 2.8} &\num{6.1\pm 0.2}	&\num{83\pm 5}		&\num{72\pm 5}\\

\hline\noalign{\smallskip}
\end{tabular}
\end{table}

One can evaluate the crystal bending angle by identifying with appropriate fits the inflection point of the error function of the losses rise (produced by the collimator that intercepts channeled particles).
The average bending angle (over several measurements) are summarised in Tab.~\ref{tab_summary}.

The normalized losses recorded during linear scans of horizontal and vertical crystals in Beam 1 and of the vertical crystal in Beam 2 both at injection and collision energies are shown in Fig.~\ref{fig_kick}.
In Fig.~\ref{sub:kick_inj}, the signals of linear scan show initial and final plateaus approximately of the same level. 
This indicates that at injection energy all crystals have comparable multiturn channeling efficiencies~\cite{prev}, and it is consistent with the observation that all crystals are far from the critical radius.
Instead, as shown in Fig.~\ref{sub:kick_ft}, crystal B1-H has a different behaviour with respect to the other two crystals at \SI{6.5}{\tera\electronvolt}.
As a matter of fact, its channeling plateau is reached well before the other crystals and its dechanneling population is enhanced when approaching small deflection angles.
The efficiencies computed for each crystal from normalized loss plateau level differences are reported in Tab.~\ref{tab_summary}.

For crystals of \SI{4}{\milli\meter} length, the critical radius is evaluated as \SI{\sim0.8}{\meter} and \SI{\sim11.7}{\meter}, for proton energy of \SI{450}{\giga\electronvolt} and \SI{6.5}{\tera\electronvolt}, respectively.
The B1 horizontal crystal, with a bending of \SI{63.2}{\micro\radian}, has a bending radius of \SI{63.3}{\meter}.
Whilst at injection energy all crystals are at least a factor \num{\sim79} above the $R_c$, at flat top the B1-H bending radius is only \num{\sim 5.4} $R_c$.
This value has to be compared to the radius of the vertical crystals that is \numlist{\sim6.1;\sim8.6} $R_c$ for B2 and B1, respectively.
As explained in Section~\ref{sec:dechanneling}, the dechanneling length in bent crystals (as the channeling efficiency \cite{biryukov}) is dependent on a bending radius: a small $L_D$ corresponds to a small bending radius. 
Thus, the dechanneling probability increases, while the channeling efficiency decreases.

An important enhancement of dechanneled population at low deflection angles in crystals close to critical radius is also observed.
This is an indication of particles that lose quickly the channeling conditions.
The main contribution is due to the interaction with nuclei, given that the extreme bending radius produces a potential well deformation that forces channeled particles to oscillate closer to the crystalline plane. 
This observation is crucial because there is no analytic description of the nuclear dechanneling process available and data from this and future work might help to solve this open point.

\section{\label{sec:simulation}Simulation Comparison}

In \texttt{SixTrack}, a Monte--Carlo routine to simulate crystal particle interactions has been developed \cite{prev,cry_rout,dan_yellow}. 
A sequence of checks applied to each particle trajectory allows to establish which cross-section formula should be applied for the random evaluation of the interaction effects. 
Such evaluation determines the outgoing trajectory of the interacting particle. 
The routine has been bench--marked with single and multi--pass data from different experiments.

\begin{SCfigure}[][t]
\includegraphics[width=0.65\textwidth]{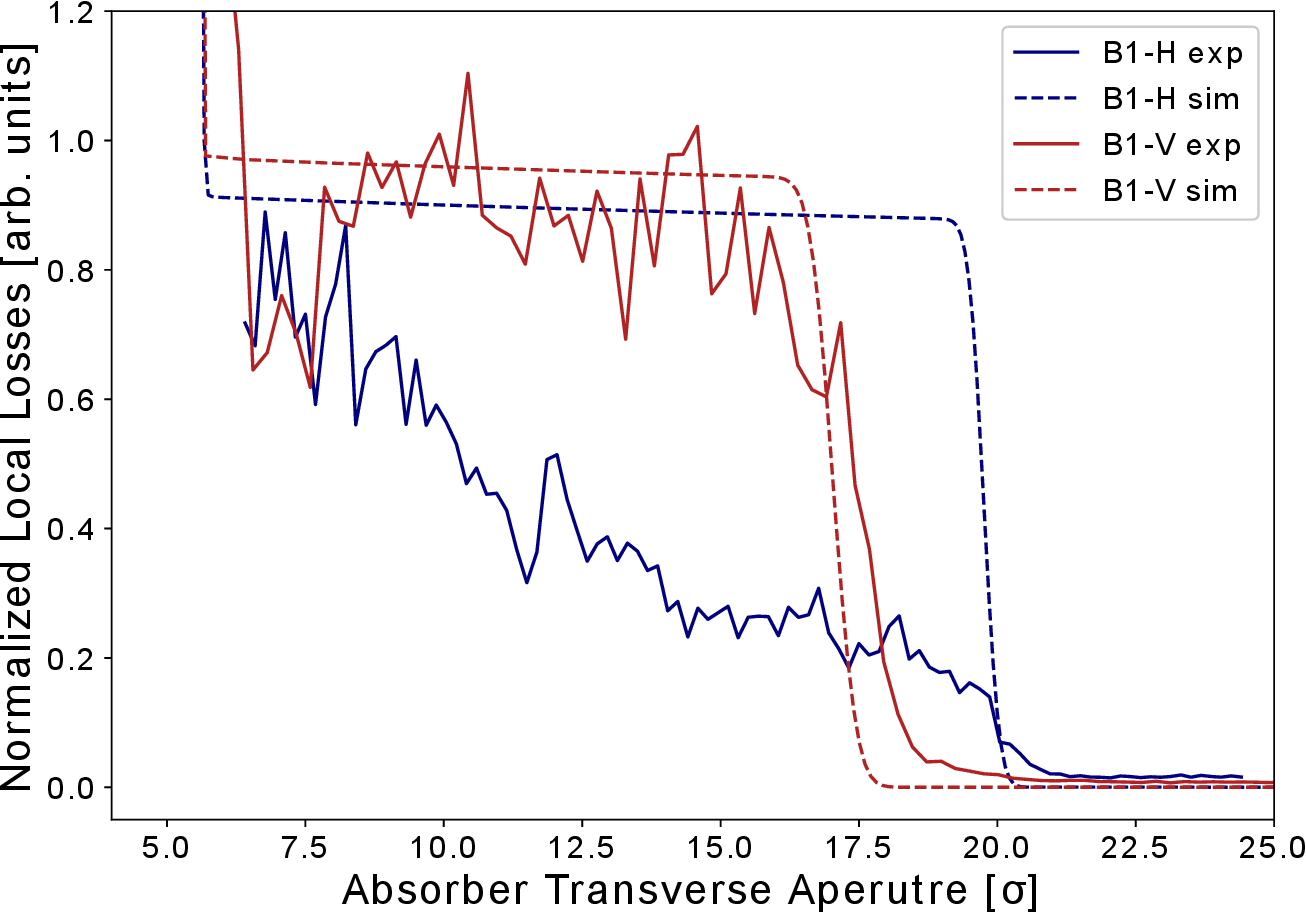}
  \caption{Beam losses normalized to beam flux and to the loss level at interception of the circulating beam as a function of jaw aperture in units of beam size [$\sigma$], during a linear scan at top energy with the crystals in channeling orientation. For B1 crystal measurements (solid lines) a comparison with simulations is presented (dashed lines). In simulations, the value bending angle values are taken from Table~\ref{tab_summary}.}
  \label{fig_sim}
\end{SCfigure}

Linear scans are simulated by sampling the particles passing through the collimators used in measurements.
The particle distributions can be integrated as a function of the collimator jaw position.
In this way, it is possible to obtain a direct confrontation between the simulation and the experimental data, as shown in Fig.~\ref{fig_sim} for the TCSG.B4L7.B1 and D4L7.B1 scans at top energy.
Here, reference is made only to B1 crystals, so as to compare the crystal that exhibits the effect under investigation with one that does not. 
Both experimental and simulated results are normalized to the values of losses and the total number of particles, respectively, at the position where the beam envelope is intercepted.

In Fig.~\ref{fig_sim}, a good agreement between data and simulation for the vertical case (red, solid and dashed lines, respectively) is evident.
The position where the channeled halo is observed is in agreement within \SI{110}{\micro\meter}, the same difference observed when the circulating beam is touched.
In Fig.~\ref{fig_sim}, the same comparison is shown for the horizontal plane (blue, solid and dashed lines, respectively). 
In this case, simulations do not reproduce correctly the data, in particular in the dechanneling region with low deflection.
The horizontal crystal has a bending angle \SI{26}{\percent} larger than specification, and with a bending radius only \num{\sim 5.4} times larger than the critical radius at LHC top energy; simulations were made using these measurements as an input.
As discussed in \cite{dan_th}, the analytic description of the nuclear dechanneling process used in the crystal simulation routine has a limited validity for crystals with a bending radius close to a few times the $R_c$. 
Because of this limitation, the routine does not reproduce well the dechanneling distribution, especially at low deflections.
Dechanneling distribution is benchmarked against data at lower energy (one order of magnitude lower), thus very far from the critical radius.
With crystals as close to the $R_c$ as the B1 horizontal, there is no benchmark available, hence the evaluation of the dechanneling population does not match observations.
The vertical crystal, with a lower deflection, has a bending radius of \SI{100}{\meter}; hence, the simulation is not affected by the same issue.

\section{\label{conclusion}Conclusions}
The LHC crystal collimation experimental layout represents a unique test bench for both collimation and crystal related studies.
In this paper, it is shown how crystals behave in presence of beams of significant energy difference.

One of the crystals is significantly more bent than the others; it has a bending radius \num{\sim 5} $R_c$ at top energy and shows a particularly enhanced dechanneling population at low deflection angle. 
This feature is also confirmed by a reduced channeling efficiency with respect to the other crystals under test.
The extreme bending, at such energy, causes a distortion too severe in the bent potential well, that reduces the efficiency of channeling and enhances a quick nuclear dechanneling of the initially trapped particles.
Compared to experimental results with other crystals, the value of \num{\sim 5} $R_c$ appears to be close to a threshold value at which nuclear dechanneling is highly intensified.

By means of simulations, were also compared the results at top energy of B1 crystals.
The crystal simulation routine has a well known limitation in reproducing dechanneling population for crystal extremely close to $R_c$ \cite{cry_rout}.
It has been shown how the simulation is able to reproduce correctly data for B1-V crystals, while it fails to duplicate the high dechanneling population at low bending for crystal B1-H.


The dependence of dechanneling on bending radius might be further investigated in a more flexible experimental environment. 
Tuning crystal bending radius to beam particle energy, it would be possible to produce an extensive study about the dechanneling dependence on it and exploit the results as input to future improvement of the simulation tools.

\section*{Acknowledgement}
The authors would like to acknowledge all the CERN teams and groups who contributed to the experimental measurements, in particular the ABP group and OP group in the Beams department and the STI and SMM groups in the Engineering department. A special acknowledgement goes to the members of the UA9 collaboration, in particular to the INFN team and the PNPI team who built the crystals installed in the LHC. This work was supported by the HL-LHC project and by the Collimation project at CERN.

\bibliographystyle{spphys}       
\bibliography{biblio_dech.bib}

\begin{thebibliography}{10}
\providecommand{\url}[1]{{#1}}
\providecommand{\urlprefix}{URL }
\expandafter\ifx\csname urlstyle\endcsname\relax
  \providecommand{\doi}[1]{DOI \discretionary{}{}{}#1}\else
  \providecommand{\doi}{DOI \discretionary{}{}{}\begingroup \urlstyle{rm}\Url}\fi

\bibitem{lindhard}
J.~Lindhard, Kongelige Danske Videnskabernes Selskab \textbf{34}, 14 (1965)

\bibitem{tsyg}
E.~Tsyganov, TM-684  (1976)

\bibitem{taratin}
A.M. Taratin, Phys. Part. Nucl. \textbf{29}, 437 (1998).
\newblock \doi{10.1134/1.953085}

\bibitem{na48doble1996novel}
N.~Doble, L.~Gatignon, P.~Grafstr{\"o}m, Nucl. Instrum. Methods Phys. Res., Sect. B \textbf{119}(1-2), 181 (1996)

\bibitem{na48fanti2007beam}
V.~Fanti, A.~Lai, D.~Marras, L.~Musa, A.~Nappi, R.~Batley, A.~Bevan, R.~Dosanjh, R.~Galik, T.~Gershon, et~al., Nucl. Instrum. Methods Phys. Res., Sect. A \textbf{574}(3), 433 (2007)

\bibitem{afonin2001high}
A.~Afonin, V.~Baranov, V.~Biryukov, M.~Breese, V.~Chepegin, Y.A. Chesnokov, V.~Guidi, Y.M. Ivanov, V.~Kotov, G.~Martinelli, et~al., Phys. Rev. Lett. \textbf{87}(9), 094802 (2001)

\bibitem{afonin2002investigations}
A.~Afonin, V.~Baranov, V.~Biryukov, V.~Chepegin, Y.~Fedotov, V.~Kotov, V.~Maisheev, V.~Terekhov, E.~Troyanov, et~al., arXiv preprint hep-ex/0207040  (2002)

\bibitem{velotti2019demonstration}
F.~Velotti, P.~Bestmann, M.~Butcher, M.~Calviani, M.~Di~Castro, M.~Donze, L.~Esposito, M.~Fraser, S.~Gilardoni, B.~Goddard, et~al., in \emph{10th Int. Particle Accelerator Conf. (IPAC'19), Melbourne, Australia, 19-24 May 2019} (JACOW Publishing, Geneva, Switzerland, 2019), pp. 3399--3403

\bibitem{lhc_ch}
W.~Scandale, G.~Arduini, M.~Butcher, F.~Cerutti, M.~Garattini, S.~Gilardoni, A.~Lechner, R.~Losito, A.~Masi, D.~Mirarchi, et~al., Physics Letters B \textbf{758}, 129 (2016)

\bibitem{redaelli2021first}
S.~Redaelli, M.~Butcher, C.~Barreto, R.~Losito, A.~Masi, D.~Mirarchi, S.~Montesano, R.~Rossi, W.~Scandale, P.S. Galvez, et~al., The European Physical Journal C \textbf{81}(2), 1 (2021)

\bibitem{dan_layout}
D.~Mirarchi, G.~Hall, S.~Redaelli, W.~Scandale, The European Physical Journal C \textbf{77}(6), 424 (2017).
\newblock \doi{10.1140/epjc/s10052-017-4985-4}

\bibitem{dan_th}
D.~Mirarchi, Vol. 31 - crystal collimation for lhc.
\newblock Ph.D. thesis, Imperial College, London (2015)

\bibitem{rossi_phd}
R.~Rossi, Experimental assessment of crystal collimation at the large hadron collider.
\newblock Ph.D. thesis, La Sapienza, University of Rome (2017).
\newblock Presented 26 Jan 2018

\bibitem{Afonin}
A.G. Afonin, V.M. Biryukov, V.A. Gavrilushkin, V.N. Gres', B.A. Zelenov, V.I. Kotov, V.A. Maisheev, A.V. Minchenko, V.N. Terekhov, E.F. Troyanov, Y.A. Chesnokov, M.G. Gordeeva, A.S. Denisov, Y.M. Ivanov, A.A. Petrunin, V.V. Skorobogatov, B.A. Chunin, Journal of Experimental and Theoretical Physics Letters \textbf{67}(10), 781 (1998).
\newblock \doi{10.1134/1.567748}

\bibitem{strip_lhc}
G.~Germogli, A.~Mazzolari, V.~Guidi, M.~Romagnoni, Nuclear Instruments and Methods in Physics Research Section B: Beam Interactions with Materials and Atoms  (2017)

\bibitem{qm_model}
Y.M. Ivanov, A.~Petrunin, V.V. Skorobogatov, JETP letters \textbf{81}(3), 99 (2005)

\bibitem{biryukov}
V.M. Biryukov, Y.~Chesnokov, V.I. Kotov, \emph{Crystal channeling and its application at high--energy accelerators} (Springer Science \& Business Media, 2013)

\bibitem{gonio2}
M.~Butcher, A.~Giustiniani, R.~Losito, A.~Masi, in \emph{IECON 2015 - 41st Annual Conference of the IEEE Industrial Electronics Society} (2015), pp. 003,887--003,892.
\newblock \doi{10.1109/IECON.2015.7392706}

\bibitem{assman}
R.W. Assmann, O.~Aberle, G.~Bellodi, A.~Bertarelli, C.~Bracco, H.~Braun, M.~Brugger, S.~Calatroni, R.~Chamizo, A.~Dallocchio, B.~Dehning, A.~Ferrari, P.~Gander, A.~Grudiev, E.B. Holzer, J.B. Jeanneret, J.M. Jiménez, M.~Jonker, Y.~Kadi, K.~Kershaw0, J.~Lendaro, J.~Lettry, R.~Losito, M.~Magistris, A.~Masi, M.~Mayer, E.~Métral, R.~Perret, C.~Rathjen, S.~Redaelli, G.~Robert-Démolaize, S.~Roesler, F.~Ruggiero, M.~Santana, P.~Sievers, M.~Sobczak, A.~Tsoulou, V.~Vlachoudis, T.~Weiler, I.S. Baishev, I.~Kurochkin.
\newblock The final collimation system for the lhc (2006)

\bibitem{rossi_note}
R.~Rossi, O.~Aberle, O.O. Andreassen, M.~Butcher, C.A. Dionisio~Barreto, A.~Masi, D.~Mirarchi, S.~Montesano, I.~Lamas~Garcia, S.~Redaelli, W.~Scandale, P.~Serrano~Galvez, A.~Rijllart, G.~Valentino, F.~Galluccio.
\newblock Beam 2 crystal characterization measurements with proton beams in the lhc (2018)

\bibitem{prev}
V.~Previtali, Performance evaluation of a crystal-enhanced collimation system for the lhc.
\newblock Ph.D. thesis, EPFL (2010)

\bibitem{cry_rout}
D.~Mirarchi, G.~Hall, S.~Redaelli, W.~Scandale, Nuclear Instruments and Methods in Physics Research Section B: Beam Interactions with Materials and Atoms \textbf{355}, 378 (2015)

\bibitem{dan_yellow}
D.~Mirarchi, S.~Redaelli, W.~Scandale.
\newblock Crystal implementation in sixtrack for proton beams (2018)

\end{thebibliography}

\end{document}